\documentclass[11pt]{article}

\usepackage{subfigure} 
\usepackage{amsmath} 
\usepackage{amssymb} 
\usepackage{graphicx} 
 
\newcommand{\negativ}[1]{\mathcal{N}{(#1)}}
\newcommand{\quoc}{{\tt quantum-octave}}
\newcommand{\qofunc}[1]{{\tt #1}}
\newcommand{\CPP}{{\tt C++}}
\newcommand{\N}{\mbox{$\mathbb{N}$}}
\newcommand{\code}[1]{{\tt #1}}

\newcommand{\ptr}[2]{\mathrm{tr}_{#1}(#2)}

\newcommand{\bra}[1]{\langle{#1}|}
\newcommand{\ket}[1]{|{#1}\rangle} 

\begin{document}
%
%
\title{Numerical simulations of mixed states quantum computation}
\author{P.~Gawron, J.~A.~Miszczk\footnote{e-mail: \{gawron, miszczak\}@iitis.gliwice.pl}\\  Institute of Pure and Applied Informatics Polish of the Academy of Science,\\ Ba\l{}tycka 5, 44-100 Gliwice, Poland}

\maketitle

\begin{abstract}
We describe \quoc{} package of functions useful for simulations of quantum algorithms and protocols. Presented package allows to perform simulations with mixed states. We present numerical implementation of important quantum mechanical operations -- partial trace and partial transpose. Those operations are used as building blocks of algorithms for analysis of entanglement and quantum error correction codes. Simulation of Shor's algorithm is presented as an~example of package capabilities.
\end{abstract}

\section{Motivation}
Most of software developed for simulations of quantum computing is based on finite dimensional Hilbert space formalism.\cite{QCL,qucalc} Mixed state model for quantum computing allows to incorporate many features which are crucial for analysis of entanglement and decoherence.

Main contribution of this work is presentation of \quoc{} package -- a tool which allows to perform simulations of quantum systems using density operators formalism in convenient way. We present algorithms used in implementation of partial operations in \quoc{}. Sample results obtained using \quoc{} present capabilities of package.
\section{Implementation in GNU Octave language}
Package \quoc{} is implemented in GNU Octave language.\cite{Eat00} List of functions and their detailed description can be found on project web page.\cite{QO04} We present here only the most important features of \quoc{}.

Most of \quoc{} functions can be applied to density operators, which are represented by normalise matrices. Package offers also functions for operations on pure states. Base ket vectors and their linear combinations are easily constructed and transformed into corresponding density operators using \qofunc{State} function, and it is easy to obtain mixtures of states using \qofunc{MixStates} function. Computation process is performed using \qofunc{Evolve(gate, state)} function, where \qofunc{gate} is unitary operation.

Lowlevel functions \qofunc{BinVec2Dec} and \qofunc{Dec2BinVec} allow for convenient use of register notation $(\ket{1}\otimes\ket{0}\otimes\ket{0}\ldots\otimes\ket{0})$. They are implemented in \CPP{} using \code{liboctave} are they are used for manipulations of registers in \qofunc{PTrace} and \qofunc{PTranspose} functions. For example sequence of operaions
\begin{verbatim}
y = Dec2BinVec (3,4);
tmp = y(2);
y(2) = y(3);
y(3) = tmp;
x = Ket(BinVec2Dec(y));
\end{verbatim}
allows to permute second and third qubits in base state in four dimensional space.

\section{Algorithms for partial operations}
Partial operations are crucial for analysis of quantum information processing. Partial transposition allows to distinguish entangled and sparable states in low dimensional systems.\cite{Hor01} This operation was used in Ref.~\cite{ViWe02} for constructing computable measure of entanglement -- negativity.\footnote{Negativity is defined as a~sum of negative eigenvalues of density matrix after partial transposition. It is easy to check that this definition is equivalent to formula $\negativ{\rho} = \frac{||\rho ^{T_A} ||_{1} - 1}{2}$, where $||\rho||_1$ denotes trace norm of density matrix $\rho$.}
%
\subsection{Partial transposition using registers}
In quantum-octave partial transposition is based on permutation of qubits indexes. Suppose we perform partial transposition with respect to some qubits in input state. Let $Q = \{q_1,q_2,\ldots,q_n\} \subset \N$ be the set of numbers labelling those qubits. Using register notation one can write
\begin{equation}
	\bra{\alpha}\rho^{T_{q_1,\ldots,q_n}}\ket{\beta} = \bra{V(\alpha,\beta,q_1,\ldots,q_n)}\rho\ket{V(\beta,\alpha,q_1,\ldots,q_n)},
\end{equation}
where
\begin{equation}
	V_i(\alpha,\beta,q_1,\ldots,q_n) = \left\{
	\begin{array}{cr}
		\beta_i, & i\in\{q_1,\ldots,q_n\}\\
		\alpha_i, & i\notin\{q_1,\ldots,q_n\}
	\end{array}
	\right.
\end{equation}
and $\alpha_i,\beta_i,V_i$ represent $i$-th number in binary representation of $\alpha,\beta$ and $V$.

\subsection{Partial trace}
For density matrix $\rho\in\mathcal{S}(\mathcal{H}_A\otimes\mathcal{H}_B)$ one can obtain description of subsytems $A$ and $B$ using partial trace operation.\cite{NiCh00,Keyl} In \quoc{} this operation is implemented as follows. Let $Q = \{q_1,q_2,\ldots,q_n\}$  be a given set of number labelling qubits. Density operator describing state of this qubits is defined as matrix with elements
\begin{equation}
		\bra{\alpha} \ptr{q_1,q_2,\ldots,q_n}{\rho} \ket{\beta} = \sum_{k_1=0}^1\ldots\sum_{k_n=0}^{1}\bra{W(\alpha,k_1,\ldots,k_n)}\rho \ket{W(\beta,k_1,\ldots,k_n)},
\end{equation}
where $\alpha$ and $\beta$ are binary digits of length $m=N-n$. Number $k_1, \ldots, k_n\in\{0,1\}$ label base vectors in Hilbert spaces of qubits $q_1, \ldots, q_n$ respectively. Numbers $W$ are constructed as follows
\begin{equation}
	W(\alpha,k_1,\ldots,k_n) = (W_1,W_2,\ldots,W_N),\ W_i = \left\{ 
									\begin{array}{cl} 
									k_i, & i\in Q\\
									\alpha_i, & i\notin Q
									\end{array}
								\right..
\end{equation}
In \quoc{} construction of numbers $W$ is performed using function \qofunc{BuildBinaryVector} implemented in \CPP{}. This is motivated by extensive usage of this function. If one wants to perform partial trace operation on $N$-qubit system with respect to its $m$-qubit subsystem, function \qofunc{BuildBinaryVector} have to be called $2^{2(N-m)N}$ times.

One should note that this algorithm allows to perform partial trace with respect to any set of qubits.
\section{Shor's algorithm with mixed sates -- fidelity and distances measures}
In this section we present sample results obtained by using \quoc{}. We analyse influence of states mixing on Shor's algorithm.\footnote{The subject of analysis is restricted to quantum part of Shor's algorithm.}
\begin{figure}[!ht]
\begin{tabular}{cc}
	\subfigure[Fidelity between density matrices as a functions of 'purity' of state]
	{
		\label{fig:fid:a}
		\includegraphics[width=0.4\textwidth, angle=-90]{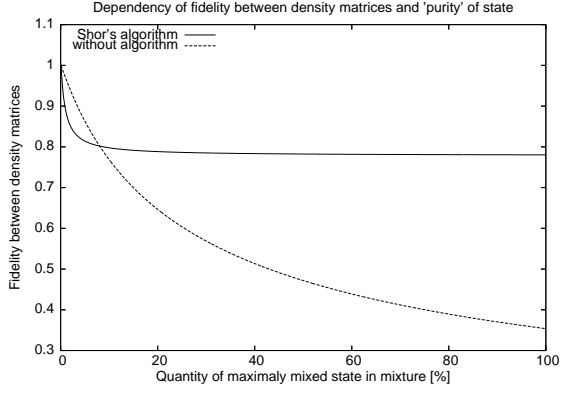}
	}
	\subfigure[Fidelity between probability distributions as a functions of 'purity' of state]
	{
		\label{fig:fid:b}
		\includegraphics[width=0.4\textwidth, angle=-90]{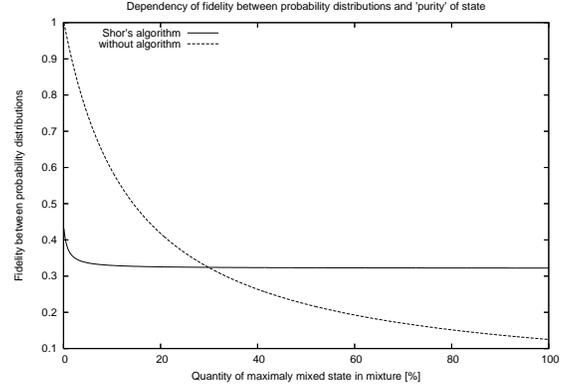} 
	}\\
	\subfigure[Trace-distance as a functions of 'purity' of state]
	{
		\label{fig:fid:c}
		\includegraphics[width=0.4\textwidth, angle=-90]{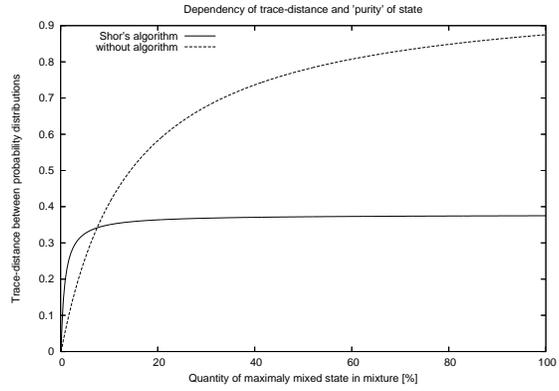}
	}
	\subfigure[Chi-square measure as a functions of 'purity' of state]
	{
		\label{fig:fid:d}
		\includegraphics[width=0.4\textwidth, angle=-90]{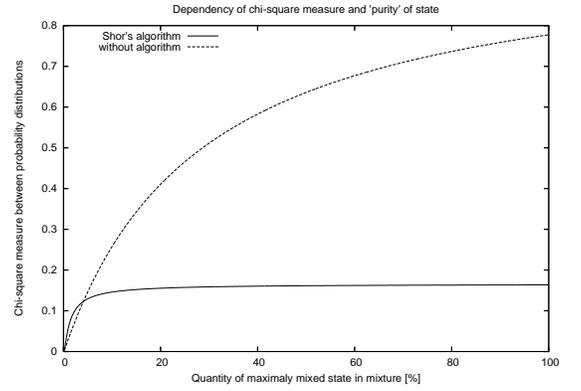}
	}
\end{tabular} 
\label{fig:main}
\caption{Comparison odistancece measures between density matrices (Fig.~\ref{fig:fid:a}) and between probability distributions (Figs.~\ref{fig:fid:b}-\ref{fig:fid:d}) for simulation of Shor's algorithm}
\end{figure}
The idea was to observe how quantum state, obtained by performing quantum algorithm on non-pure state, differs from state obtained in ideal (pure state) case. On each step of simulation Initial state $\rho_p$ was prepared as mixture of pure state $\rho_0=\vert \underbrace{00\ldots 0}_d \rangle \langle \underbrace{00\ldots 0}_d \vert$ and maximally mixed state $I_d/d$ 
\begin{equation}
\rho_p=(1-p)\rho_0+pI_d/d,
\end{equation}
where $p\in\{0.00,0.01,0.02, \ldots, 1.00\}$ and $I_d$ is $d$-dimensional identity matrix. For each state obtained in such way the trivial and Shor's circuits were applied and for each obtained outcome two distances and two fidelity measures were calculated. Results of those simulations are presented in Fig.~1. Shor's algorithm is performed on 7 qubits, but only the first 3 qubits are taken into account during calculations of fidelity and distance measures because the last 4 qubits decohere and normally aren't measured.

Fidelity between density matrices, as given in Ref.~\ref{equ:fid} is proper measure which shows how much two mixed states differ.\cite{NiCh00}
To compare different methods of calculating distance between states we have chosen three functions (Eqs.~\ref{equ:probfid}, \ref{equ:chi2} and~\ref{equ:trdist}) that operate on probability distributions. Probability distributions are obtained by performing measurement of observable $\widehat{Z}\otimes\widehat{Z}\otimes\widehat{Z}$. Formulas used to calculate measures are presented below. Let $p_1,p_2$ be probability distributions, and $\rho_1, \rho_2$ -- density matrices.
\begin{itemize}
\item Fidelity beetween density matrices
\begin{equation}
	F(\rho_1, \rho_2)=tr\sqrt{\sqrt{\rho_1}\rho_2\sqrt{\rho_1}},
	\label{equ:fid}
\end{equation}
\item Fidelity beetwen probability distributions
\begin{equation}
	F(p_1,p_2)=\sum_{x\in X}\sqrt{p_1(x)}\sqrt{p_2(x)}, 
	\label{equ:probfid}
\end{equation}
where  $p=\frac{1}{2}(p_1+p_2);$
\item $\chi^2$-measure
\begin{equation}
	\chi^2(p_1,p_2)=\sum_{x\in X}\frac{(p_1(x)-p(x))^2}{p(x)},
	\label{equ:chi2}
\end{equation}
\item Trace-distance
\begin{equation}
	F(p_1,p_2)=\frac{1}{2}\sum_{x\in X}\vert{p_1(x)-p_2(x)}\vert.
	\label{equ:trdist}
\end{equation}
\end{itemize}
One can conclude that for small addition of noise results differs wery much from the ideal case. The loss of quality grows slower for bigger contribution of noise.
\section{Final remarks}
Package \quoc{} allows to perform simulations of mixed states quantum computation in convenient way. It also provides functions for analysis of entanglement and quantum errors. Quantum protocols and algorithm such as teleportation,\cite{Tele} qubit authentication\cite{Curty} or Shor's algorithm\cite{Shor} can be easily implemented as GNU Octave functions and the use in futher simulations. This allows to perform simulations operating on high level quantum primitives (e.g. qubit authentication, error correcting code). 
 
\end{document}